\documentclass[journal=jacsat,manuscript=article]{achemso}
\usepackage{color,soul}
\soulregister\cite7
\soulregister\ref7
\soulregister\pageref7
\usepackage{xcolor}
\usepackage[english]{babel}
\usepackage{wrapfig}
\usepackage{hyperref}
\usepackage{breqn}
\usepackage{amsmath}
\usepackage{siunitx}
\usepackage[version=4]{mhchem}
\usepackage{graphicx} 
\usepackage{color}
\usepackage{float} 
\DeclareSIUnit\angstrom{\text{Å}}
\DeclareSIUnit\rpm{rpm}

\usepackage{subcaption}
\DeclareSIUnit\ppm{ppm}

\title{Morphology control and low-temperature magnetotransport in chiral 2D perovskite R-(MBA)$_2$PbI$_4$}

\author{ \textsuperscript{$\dagger$}Shehreen Aslam, \textsuperscript{$\dagger$}Sam Saiter, Bradley Lloyd, Paul Kliewer, Matthew P. Hautzinger, Matthew C. Beard, and Meenakshi Singh}
\email{msingh@mines.edu}
\begin{document}

\begin{abstract}
\noindent Two-dimensional chiral hybrid perovskites, such as R/S-(MBA)\textsubscript{2}PbI\textsubscript{4}, are leading candidates  for realizing and studying chirality-dependent charge and spin transport. However, their prohibitive in-plane resistance has precluded the  electrical characterization. Here, we overcome this bottleneck by engineering the thin-film morphology of the chiral perovskite $R\text{-(MBA)}_2\text{PbI}_4$, enabling the first robust lateral device integration. In Hall-bar geometries, we demonstrate  Hall measurements under dark conditions, unambiguously identifying p-type conduction with a Hall mobility of $\sim 0.2 \text{cm}^2    \text{V}^{-1} \text{s}^{-1}$ and a carrier density of $\sim 3\times10^{14} \text{ cm}^{-2}$, parameters previously inaccessible in this class of materials. Furthermore, we observe enhanced magnetoresistance along transport paths crossing grain boundaries, highlighting the strong influence of morphology on  in-plane transport. This work demonstrates  in-plane magnetotransport, enabling future
investigations of the fundamental mechanisms of chirality-induced spin selectivity (CISS)
and accelerating the integration of chiral materials into functional spintronic devices.

 \end{abstract}

\vspace{0.02 in}

\section{Introduction}

Chiral materials have attracted significant attention due to their ability to control electronic spin without an applied magnetic field, which opens new opportunities in the field of spintronics, quantum information, and optoelectronics~\cite{yang2021chiral,foo2025mind,wang2020spin}. The mechanism behind this functionality is  chirality-induced spin selectivity (CISS), which is believed to spin polarize electrons as they are transmitted through a chiral medium~\cite{r4,r5}. Strong evidence of CISS has been reported in various systems, including biomolecules such as DNA and peptides, as well as synthetic frameworks such as helicenes~\cite{r6,r7,r8}. However, systematic, quantitative transport measurements on these systems remain limited as their integration into multi-terminal devices is quite challenging. The resulting scarcity of  experimental evidence from such measurements is  a key reason why the mechanism underlying CISS remains a matter of debate~\cite{evers2022theory}. The two-dimensional (2D) chiral hybrid perovskite R-(MBA)$_2$PbI$_4$ [MBA = methylbenzylammonium] offers a promising path to solving this problem by providing a versatile platform for device integration and multi-terminal magnetotransport measurements, as it is semiconducting and can be processed into uniform thin films. \\
\vspace{0.02 in}

\noindent In 2D hybrid organic–inorganic perovskites (HOIPs), molecular chirality is typically introduced by the organic spacer cation, while the layered inorganic framework provides strong spin–orbit coupling. In particular, in R/S-(MBA)$_2$PbI$_4$ the chiral MBA cation induces asymmetry within the inorganic layers. HOIPs have been widely studied for their photovoltaic properties. Transport studies of these systems have thus focused mainly on vertical device geometries such as spin valves and spin light-emitting diodes~\cite{lu2019spin,r12}. Researchers have demonstrated various spintronic functionalities in these out-of-plane devices, including spin-filtering response, circularly polarized electroluminescence, and magnetoresistance (MR), all driven by the CISS effect~\cite{r10,r13}. In contrast, lateral device structures, particularly Hall-bar configurations, remain relatively unexplored. These devices typically exhibit very high in-plane resistances, often in the G$\Omega$ range, which makes measurements especially challenging. To avoid this issue, most  reported studies have relied on the photo-Hall effect, which utilizes illumination to generate additional charge carriers and thus reduces resistance~\cite{Chen2016}. However, both from an electron-spin interaction perspective and a spintronics perspective, it is essential to characterize the  in-plane transport properties of the material. These  in-plane transport measurements are the focus of this study.\\
\vspace{0.02 in}

\noindent Charge transport in 2D hybrid perovskites is strongly influenced by film quality. Large grain sizes,
fewer grain boundaries, and the absence of pinholes result in more uniform films with improved film conductivity. Most efforts to enhance charge transport have been driven by
solar-cell applications, where researchers have employed process-, additive-, and cation-engineering
strategies to optimize morphology and crystallinity for improved device efficiency.
Process-engineering approaches, such as solvent-ratio control and hot casting, have been
demonstrated to enhance crystallinity and grain size, thus facilitating improved carrier
percolation~\cite{r18,r19}. Additive engineering, using compounds such as \ce{NH\textsubscript{4}SCN},
\ce{NH\textsubscript{4}Cl}, \ce{MACl}, and {\ce PbI\textsubscript{2}} has been employed to tune film orientation and
reduce trap states~\cite{r20}. Similarly, cation engineering both with small ions (\ce{FA+},
\ce{MA+}, \ce{Cs+}) and bulky organic spacers (PEA, F-PEA, MeO-PEA) has been used to
adjust interlayer coupling and reduce grain-boundary scattering~\cite{r21,r22}. More recently,
spin-speed studies have shown that varying the spin-coating speed directly tunes film thickness,
which in turn modulates morphology and grain size, thereby impacting conductivity~\cite{r23}. While these methods have been optimized for photovoltaic performance, the impact of
morphological control on  charge transport has not been studied.\\
\vspace{0.02 in}

\noindent In this work, a thickness-tuning strategy was utilized by adjusting the spin-coating speed to obtain R-(MBA)\textsubscript{2}PbI{\textsubscript{4}} films with reduced resistance suitable for integration into lateral transport studies. Subsequently, we performed comprehensive measurements on
three-terminal devices in FET geometry, examining current–voltage response, gate modulation, temperature-dependent resistance down to 38~K, and MR. Finally, we
advanced to Hall-bar geometries, where we measured Hall and temperature dependent MR effects in the dark. These results overcome the
long-standing challenge of achieving measurable in-plane transport, extract critical transport parameters using Hall measurements, and establish the influence of grain-boundaries on in-plane transport, laying the groundwork for future integration into multi-terminal devices. 
\section{Experimental Details}
\subsection{Sample Preparation}

Highly doped Si/SiO$_2$ wafers with an oxide thickness of 1~$\mu$m and a resistivity of \(1\text{-}10~\Omega\cdot\text{cm}\) (University Wafer) were used as substrates. The wafers were cleaned by sequential ultrasonication in acetone, methanol, and isopropanol for 10 minutes each, and dried under a nitrogen stream. Patterning into Hall-bar and three-terminal (FET-type) device geometries was subsequently carried out using a maskless laser-writer lithography system (Nanyte Beam UV). For our FET-type devices, relatively large contacts were used to reduce channel resistance, with drain and source electrodes 500~$\mu$m wide separated by a 5~$\mu$m channel.  For Hall-bar devices, 2~$\mu$m wide current electrodes were separated by a 10~$\mu$m channel, and the voltage-probe spacing was 2~$\mu$m. Gold contacts have been shown to diffuse into and chemically react with halide perovskites, potentially damaging the perovskite film at the interface\cite{Lai2023}.Therefore, we employed platinum contacts, which provide significantly improved stability, using a 40~nm Pt layer with a 10~nm Ti adhesion layer for all devices. Contacts were deposited by electron-beam evaporation carried out at a base pressure of $7.5 \times 10^{-8}$~Torr and a rate of 0.5~\AA~s$^{-1}$. An aluminum (Al) layer deposited on
the back of the wafer served as the back gate, with the SiO$_2$ layer acting as the dielectric. The patterned device was treated with UV--ozone at 50~$^\circ$C for 10~min prior to spin coating of the CISS film to improve surface hydrophilicity and film adhesion. Chiral R-(MBA)$_2$PbI$_4$ crystals were synthesized following a reported protocol
\cite{lu2019spin}, and their phase purity was confirmed by X-ray diffraction (XRD) before use. Briefly, lead (II) iodide (413 mg, 0.895 mmol; Sigma Aldrich) was dissolved in 5.5 mL of hydroiodic acid (Beantown chemical) and 0.5 mL of hypophosphourous acid with gentle heating. Subsequently,  (R)-(-)-$\alpha$ -methylbenzylamine (228   $\mu{L}$ , 1.80 mmol; Sigma Aldrich) was added to the solution. The mixture was heated to 100$^\circ$ C until completely dissolved, then allowed to cool to room temperature at a rate of -1$^\circ$  per minute. The orange needle-like crystals were collected with vacuum filtration, rinsed with diethyl ether (Oakwood Chemical) and dried overnight under vacuum. To prepare the
precursor solution, the synthesized crystals were dissolved in anhydrous DMF (DMF,anhydrous, 99.8\%, Sigma-Aldrich) to yield a final concentration of 50~mg in 0.25~mL, stirred until complete dissolution was observed, and freshly prepared for each deposition to minimize degradation. The precursor solutions were then spin-coated on top of the
pre-patterned and UV-ozone treated substrates inside a nitrogen-filled glovebox (H$_2$O and O$_2$ levels $< 0.1$~ppm) at spin speeds of 1000, 1500, 2000, 3000, and 4000~rpm using a programmable spin coater (WS-400B-6NPP/LITE). Immediately after deposition, the films
were annealed on a hot plate at 70~$^\circ$C for 25~min to promote crystallization, and the resulting thicknesses were measured using a stylus profilometer (D-600).

\subsection{Characterization Techniques}

X-ray diffraction (XRD) of the R-(MBA)$_2$PbI$_4$ thin films was performed using a
Panalytical Empyrean X-ray diffractometer equipped with Cu K$\alpha$ radiation
($\lambda = \SI{1.5406}{\angstrom}$), operated at \SI{45}{kV} and \SI{40}{mA}. Scans were
conducted over the $2\theta$ range of \SIrange{5}{50}{\degree} with a step size of
\SI{0.02}{\degree}. Surface morphology of the thin films was investigated using scanning
electron microscopy (FEI Helios Nanolab 600I) at an accelerating voltage of \SI{10}{kV}.\\
\vspace{0.02 in}

\noindent Magnetotransport characterization was performed using an AttoDRY 2100 closed-cycle cryostat equipped with a superconducting magnet and variable-temperature control in the range of \SIrange{1.5}{300}{K}. The measurements were conducted using standard FET and Hall-bar geometries, with the magnetic field applied perpendicular to the film plane. A Keithley 6221 current source was used to
supply a constant current, while the resulting voltages (both longitudinal, $V_{xx}$ and transverse Hall, $V_{xy}$) were measured using either a Keithley 616 electrometer or a
Keithley 2182 nanovoltmeter. The electrometers were useful for this measurement, as their high input impedance ($> \SI{100}{T\ohm}$) far exceeds the resistance of the device under test, even at low temperatures. At the same time, their noise levels ($\sim \SI{10}{\micro\volt}$) are low enough to resolve small signals.

\section{3. Results and Discussion}

\subsection{3.1 Spin-Speed Dependent Structural, Morphological, and Electrical characterization of R-(MBA)$_2$PbI$_4$ Films}

\begin{figure}[H]
  \centering
  \includegraphics[width=0.9\linewidth]{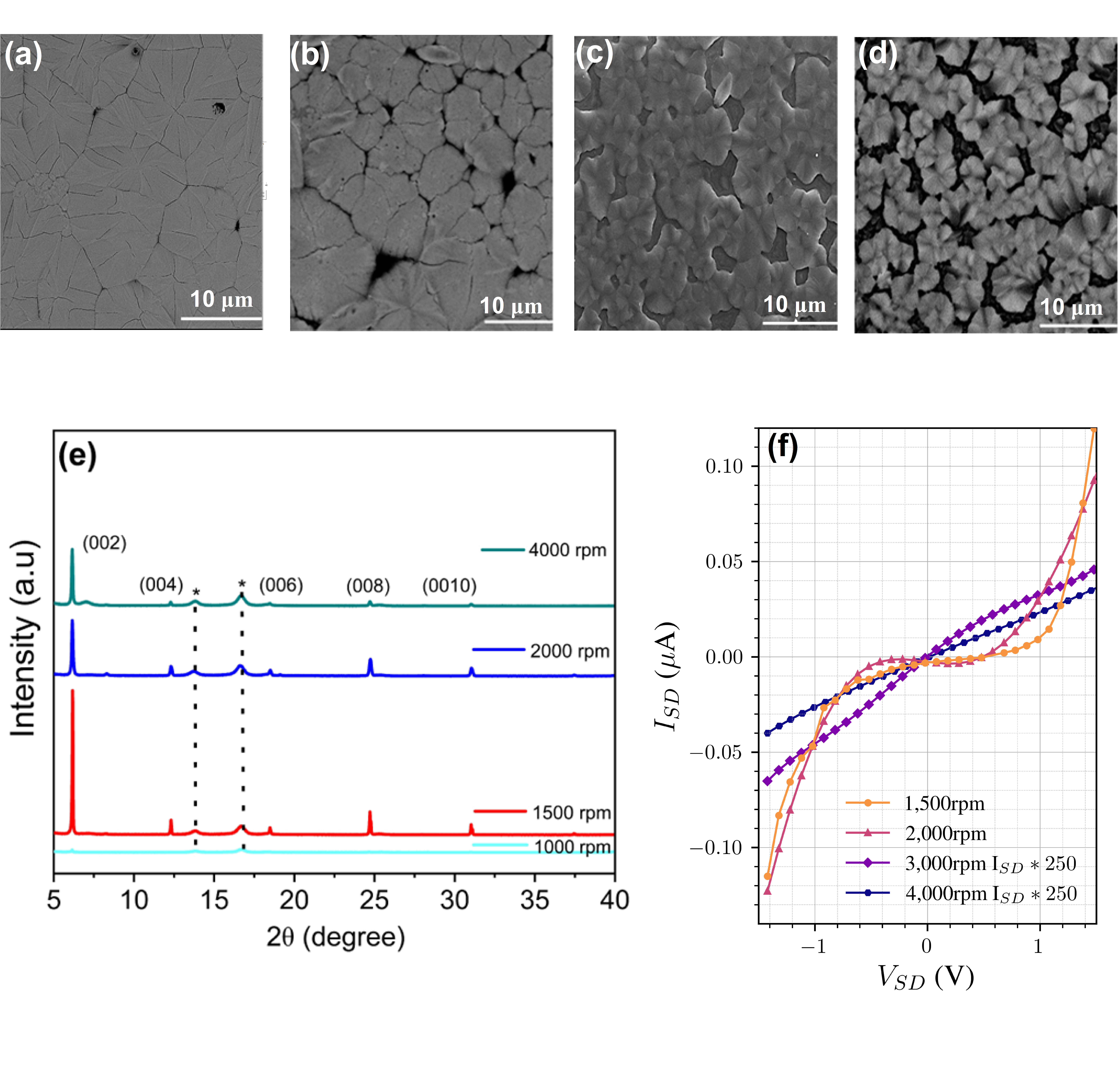}
\caption{Spin-speed dependence of the morphology, structure, and electrical properties of
\((\mathrm{R\!-\!MBA})_{2}\mathrm{PbI}_{4}\) thin films. (a–d) scanning electron microscope (SEM) images of films deposited at
1500, 2000, 3000, and 4000~rpm, showing that lower spin speeds result in larger grain size,
higher uniformity, and fewer pinholes. (e) XRD patterns, with films spun above 1000~rpm
showing clear (002) reflections, with the 1500~rpm film giving the highest intensity. (f)
Current ($I_{SD}$) vs voltage ($V_{SD}$) between source and drain of the FET device,
highlighting the influence of film thickness on electrical behavior. $I_{SD}$ was multiplied
by a factor of 250 for spin speeds of 3000~rpm and 4000~rpm for legibility, as the $I_{SD}$
values for these spin speeds are in the nA range.}
    \label{fig:spin_speed_dependence}
\end{figure}

\noindent A series of R-(MBA)$_2$PbI$_4$ films was prepared to examine how processing conditions influence their morphology, crystallinity, and electrical response. The spin-coating speed was varied from 1000 to 4000~rpm to generate films of different thicknesses. Structural characterization was performed using SEM and XRD to assess film morphology and
crystallographic quality. The corresponding electrical behavior was evaluated using three-terminal FET devices, which provided the current--voltage characteristics for the different film thicknesses (Figure~\ref{fig:spin_speed_dependence}). The contact
resistance extracted from multi-terminal structures averaged $\sim 50~\text{k}\Omega$ and was negligible compared to the film resistance.\\
\vspace{0.02in}

\noindent The film morphology was found to be highly dependent on spin speed, governed by the balance between fluid flow and solvent evaporation.\cite{bornside1989spin} At 1000 rpm, the centrifugal force is insufficient to overcome the liquid surface energy, resulting in incomplete spreading of the precursor across the substrate
and the formation of pooled regions, which causes significant non-uniformity\cite{kotsuki2014importance}. This non-
uniformity leads to uncontrolled nucleation and crystal growth, producing isolated grains with poor inter-grain connectivity. As a result, a continuous film is not formed, preventing the establishment of effective lateral conduction pathways. Consequently, despite the larger
thickness, the film exhibits extremely high resistance beyond the measurable limit, and this condition was excluded from further analysis.  At \SI{1500}{rpm}, the resulting film had an average thickness of $\sim\!300~\text{nm}$
(Figure~\ref{fig:spin_speed_dependence}a) and exhibited large, well-connected grains that formed a continuous layer with minimal pinholes and voids. This morphology correlated with the lowest measured resistance ($\sim\!\SI{10}{M\ohm}$)
(Figure~\ref{fig:spin_speed_dependence}f). Increasing the spin speed to \SI{2000}{rpm} (Figure~\ref{fig:spin_speed_dependence}b) produced thinner films ($\sim\!225~\text{nm}$) with slightly smaller grains and reduced domain connectivity. Although the resistance
remained in the M$\Omega$ range, it increased modestly compared to \SI{1500}{rpm}. The trend became more pronounced at \SI{3000}{rpm} ($\sim\!165~\text{nm}$) and \SI{4000}{rpm}
($\sim\!100~\text{nm}$) (Figures~\ref{fig:spin_speed_dependence}c and d), where the films consisted of very small grains with visible gaps and poor connectivity, leading to the highest resistance values ($\sim\!\text{G}\Omega$). Electrical breakdown was also measured for these samples, establishing a rough relationship between breakdown voltage and film thickness (Supplementary Fig. S3). This pronounced increase in resistance with spin speed cannot be explained by thickness alone. Although part of the resistance reduction at lower spin speeds is expected simply from the larger film thickness, this geometric effect alone cannot account for the full trend. Because resistance is inversely proportional to film thickness, thickness alone would predict only an approximately threefold increase in resistance between the \SI{1500}{rpm} and \SI{4000}{rpm} films if the  resistivity were unchanged. Assuming similar contact resistances, the much larger observed increase ($R_{4000\mathrm{rpm}}/R_{\mathrm{1500rpm}}\approx850\,$ at 1.5V) indicates that higher spin speeds films with smaller grains and a higher density
of grain boundaries, which act as dominant scattering centers and hinder charge transport.\\
\vspace{0.02 in}

\noindent XRD patterns of R-(MBA)$_2$PbI$_4$ films prepared at different spin speeds are shown in
(Figure~\ref{fig:spin_speed_dependence}e). Films prepared at spin speeds above \SI{1000}{rpm} exhibit a strong diffraction peak, in agreement with previous reports
\cite{lu2019spin}. \\
\vspace{0.02in}
The position of the main (002) diffraction peak remains unchanged across all spin speeds, confirming that the crystalline phase is not affected by this processing
parameter. A slight enhancement in the (002) peak intensity is observed for the \SI{1500}{rpm} film, consistent with its improved film uniformity. In addition, weak reflection peaks were observed in all samples and remained unchanged with spin speed. These peaks may be related to a minor intermediate phase, but they are too weak to affect the R-(MBA)$_{2}$PbI$_{4}$ structure.
 The film deposited at \SI{1000}{rpm} shows
very weak and poorly resolved (002) reflections. Although this film appears thicker due to solution pooling at low spin speed, the slow solvent evaporation and incomplete annealing of the thick region reduce structural ordering and result in a low coherent diffracting volume, leading to weak peaks. Overall, these results confirm that spin speed does not alter the underlying phase of the R-(MBA)$_2$PbI$_4$ film, but it has a strong influence on film quality, with the \SI{1500}{rpm} film showing the highest degree of order.
 
\noindent Taken together, these results establish a clear correlation between thickness, morphology, and resistance: lower spin speeds yield thicker films with larger grains that facilitate lateral conduction, whereas excessively slow spin speeds at \SI{1000}{rpm} produce non-uniform, poorly annealed films that degrade transport despite their thickness. The \SI{1500}{rpm} condition thus represents an optimal regime in which both morphology and electrical performance are most favorable for lateral, multi-terminal device fabrication.

 \subsection{3.2 Gate-Tunable and Temperature-Dependent Longitudinal Transport}

\begin{figure}[t]
  \centering
  \includegraphics[
    width=\linewidth,
    height=0.62\textheight,
    keepaspectratio
  ]{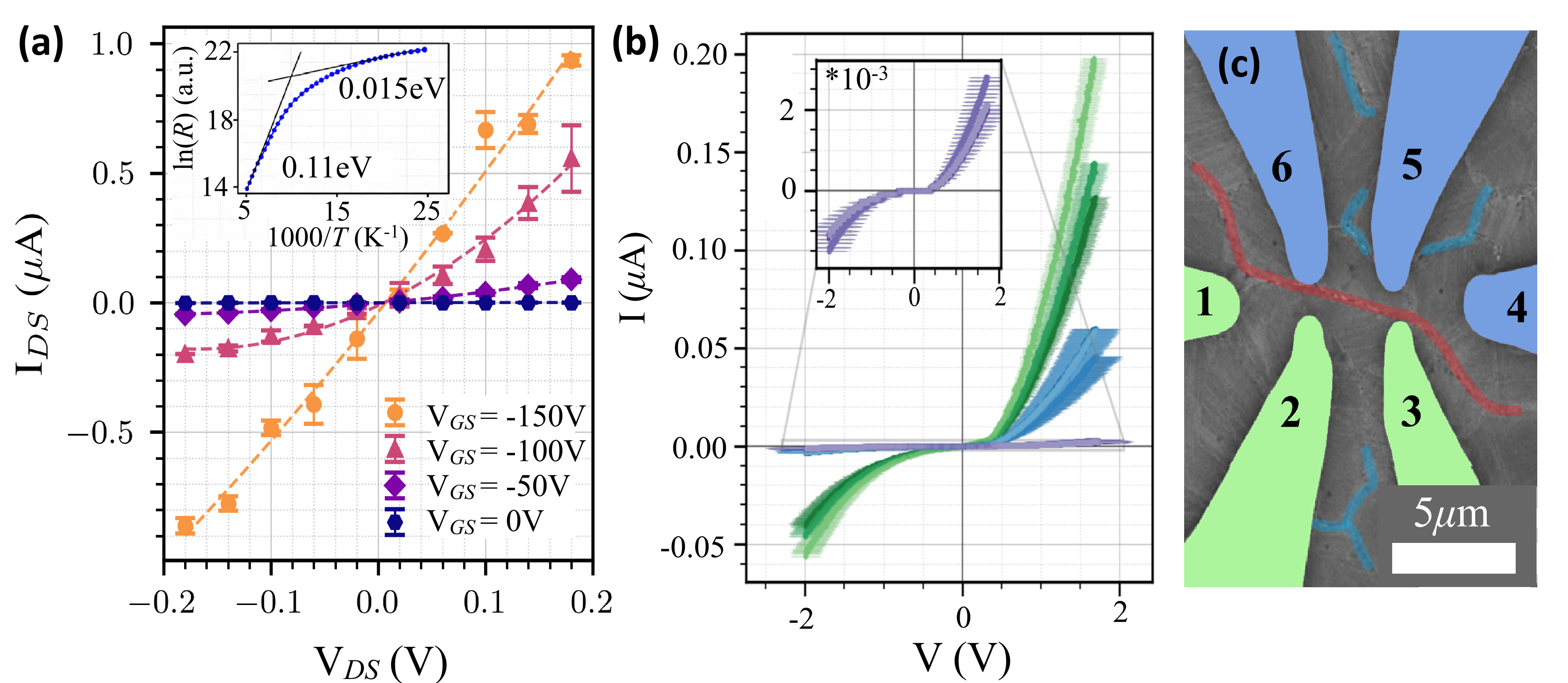}

  \caption{(a) Output characteristics of an R-(MBA)$2$PbI$4$ film spun at 1500 rpm, showing source--drain current $I_{DS}$ as a function of source--drain voltage $V_{DS}$ for several gate--source voltages $V_{GS}$. The strong gate response for negative bias is consistent with p-type conduction. Inset: Temperature-dependent resistivity of a morphology-optimized film, plotted as $\ln(R)$ versus $1000/T$, showing two Arrhenius regimes with activation energies of 0.11 eV for $T > 140$ K and 0.015 eV for $T < 70$ K. (b) Current--voltage ($I$--$V$) measurements of the Hall-bar device for different current paths. Traces measured between contacts on the same side of the grain boundary are color-coded by region: green curves correspond to contacts on one side of the grain boundary, and blue contacts the other side. Measurements spanning the grain boundary are shown in purple. Inset: Expanded view of the purple traces, included to highlight the higher-resistance current paths spanning the grain boundary. (c) Micrograph of the Hall-bar device. A grain boundary traversing the center of the Hall bar (red) separates the device into two regions. Non-contributing grain boundaries that are shunted by the metallic contacts are also highlighted (blue). Contacts on one side of the primary (red) grain boundary are shown in green, while contacts on the opposite side are shown in blue. }
  \label{fig2}
\end{figure}

\noindent The source–drain current ($I_{SD}$) versus source–drain voltage ($V_{SD}$) behavior at zero
gate voltage ($V_{GS} = 0$) was measured for the FET devices
(Figure~\ref{fig:spin_speed_dependence}f). The S-shaped current–voltage characteristics are indicative of charge transport limited by Schottky barriers at the metal–semiconductor contacts. The film
with the lowest room-temperature resistance was measured over a range of gate voltages
(Figure~\ref{fig2}a). Large gate voltages were required due to the thick
\SI{1}{\micro\meter} SiO$_2$ dielectric layer, and a
significant response was only observed for $V_{GS} < 0$. The decreasing resistivity with
increasing negative $V_{GS}$ suggests p-type semiconducting behavior with holes, as the
majority carriers. To assess possible ion-migration-induced hysteresis, the gate response was checked using both forward and reverse $V_{GS}$ sweeps. No significant hysteresis was observed, confirming that at the time-scales at which these measurements were performed, the contribution of ion-migration is negligible. Enhanced conduction at negative gate bias was consistent in both sweep directions, supporting the assignment of p-type transport.\\
\vspace{0.02in}

\vspace{0.02in}
\noindent In-plane resistance measurements were performed down to \SI{38} {K}
(Figure~\ref{fig2}a, inset). Because leakage current increases with applied bias, all measurements were performed at low excitation voltages to minimize
leakage. Using a 200~G$\Omega$ test load under the same measurement conditions, the
leakage current of our setup was verified to be $<\!10$~pA for the voltages applied to the
films. At $T < \SI{38}{K}$, the sample resistance exceeds 200~G$\Omega$, approaching the
limits of our measurement setup; reliable resistance measurements could therefore not be
obtained at lower temperatures. Even so, this measurement extends the low temperature limit of electrical characterization for this material, as in prior temperature-dependent chiral-HOIP FET studies measurements were reported down to 77–78~K \cite{Hu2024,Hu2023}.  A clear transition between different Arrhenius transport
regimes is present, for which the activation energy can be determined from the relation
$R(T) = R_0 \exp\!\left(E_a / (k_{\mathrm{B}}T)\right)$. Analysis reveals two regimes: at high
temperatures ($T > \SI{140}{K}$), the slope yields $E_a \approx \SI{0.11}{eV}$, whereas at
low temperatures ($T < \SI{70}{K}$), $E_a \approx \SI{0.015}{eV}$ is extracted. We
attribute the \SI{0.11}{eV} barrier to thermal activation of carriers within
R-(MBA)$_2$PbI$_4$. In contrast, the small low-temperature activation energy in a
few meV range is consistent with transport limited by shallow traps or hopping within an
impurity band. Low-temperature transport in some 2D semiconductors has been described within a variable-range hopping (VRH) framework, where linear behavior is expected in plots of $\ln(R)$ versus $T^{-1/3}$ for 2D VRH or $T^{-1/4}$ for 3D VRH. Here, however, the corresponding fits did not conclusively distinguish a VRH model from an Arrhenius-type transport mechanism (Figure~S2). 
\\
\vspace{0.02 in}

\noindent There is limited existing literature, both theoretical and experimental, on
transport in R-(MBA)$_2$PbI$_4$. The only comparable experimental study \cite{Hu2024} measured resistivity down to \SI{78}{K} and reported qualitatively similar temperature-dependent behavior. However, the films investigated in that study were quasi-2D mixed-$n$ perovskites, whereas the present work focuses on a strictly two-dimensional ($n=1$) R-(MBA)$_2$PbI$_4$ lattice. Differences in dimensionality and interlayer coupling naturally influence the apparent activation energies. The low-temperature activation energy reported in Ref.[22] is indeed slightly larger than the $\SI{0.015}{eV}$ measured here. In our data, the resistivity exhibits a broad crossover between the high-temperature freeze-out regime and the low-temperature shallow-trap (or impurity-band) transport regime, so the apparent slope of $\ln R$ versus $1/T$ depends sensitively on the temperature window used for the fit. Because our measurements extend to lower temperatures in a structurally well-defined $n=1$ system, the fit can be restricted to the asymptotic low-temperature regime, where the slope has flattened, yielding a smaller activation energy. By contrast, in quasi-2D mixed-$n$ films, where interlayer coupling and phase heterogeneity may broaden the crossover region, a fit performed over a higher-temperature range that still overlaps the crossover will be biased by residual freeze-out contributions, leading to a somewhat larger effective activation energy even when the underlying mechanism is similar.

\noindent In addition to the FET devices, thin film samples were prepared prepared with Hall-bar patterned contacts. Room-temperature I-V measurements were performed on a morphology-optimized film spun at 1500 rpm, and all measurements were conducted under dark conditions. To compare with the transport measured on the FET samples, 2W I-V measurements were performed between all Hall-bar contacts (Figure 2b). Interestingly, current paths between certain pairs of contacts showed resistances orders of magnitude larger than others. Three and four-wire measurements showed that contact resistance never exceeded $5\%$ for any current path, indicating that this anisotropy originates in the film. SEM imaging of the measured device reveals a pronounced grain boundary extending along the Hall-bar channel (Fig. 2c). Current paths traversing this grain boundary exhibit substantially higher resistance. Although several smaller, secondary grain boundaries are visible within the device area, their effect on the measured transport appears to be much weaker than that of the prominent boundary crossing the Hall-bar channel. As highlighted in figure 2c, many of these secondary boundaries are spanned by the metallic contacts, which likely provides low-impedance pathways that shunt their contribution to the measured resistance. In contrast, the central grain boundary lies directly in the transport path of many of the Hall-bar current configurations, forcing current to cross the boundary and producing the much larger resistance observed for those paths. At 1V, paths avoiding the grain boundary showed resistances of approximately 10--50~M$\Omega$, whereas paths crossing the boundary showed resistances of approximately 1.5~G$\Omega$, corresponding to an increase of roughly two orders of magnitude. Possible reasons for this are explored after the discussion of the MR measurements.
\vspace{0.02 in}
\\

\subsection{3.3 Temperature Dependent Magnetoresistance (MR) Measurements}
\begin{figure}[h]
  \centering
  \includegraphics[width=\textwidth]{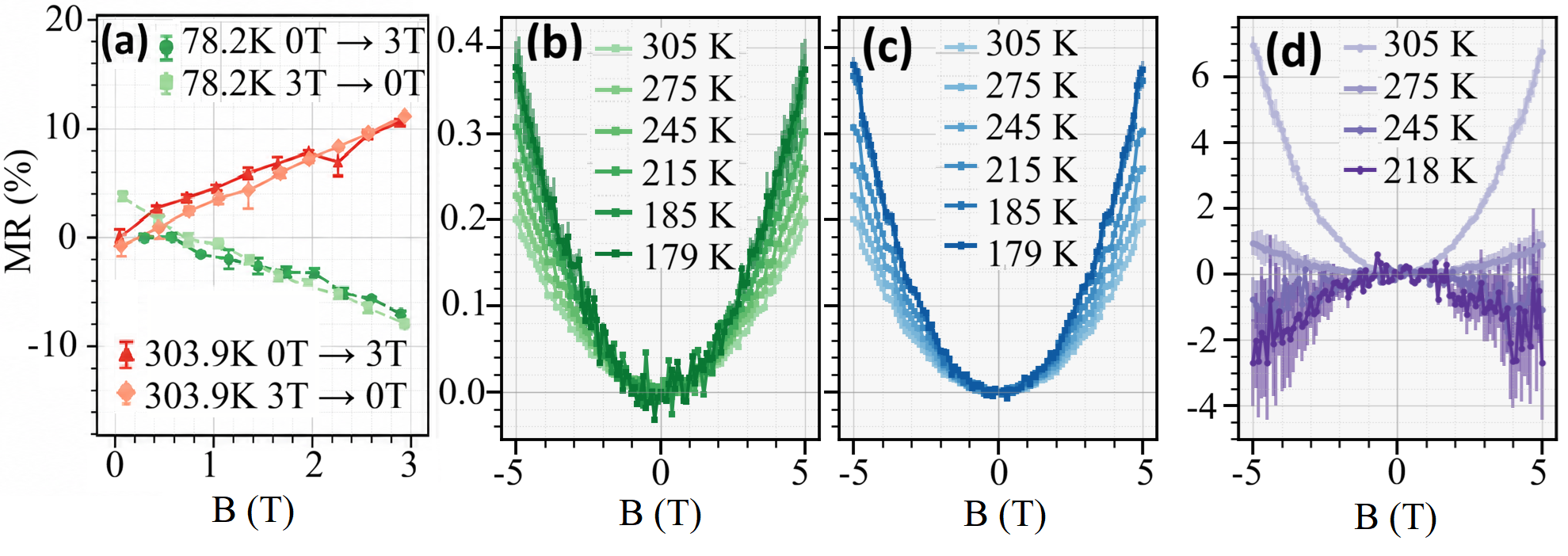}
  \caption{(a) Magnetoresistance (MR) measured in the three-terminal FET geometry at 303.9 K and 78.2 K. The room-temperature response is positive and increases monotonically with magnetic field, whereas the low-temperature response becomes negative, indicating a change in the dominant transport mechanism upon cooling. Forward and reverse field sweeps show minimal hysteresis. (b--d) Temperature-dependent Hall-bar MR measured for different current paths across the same device. In this geometry, current paths were grouped according to their relation to the grain-boundary-defined paths: green traces correspond to paths confined to one side of the boundary, blue traces correspond to paths confined to the other side, and purple traces correspond to paths spanning the grain boundary. Averaged MR responses are shown for paths within (b) green paths, (c) blue paths, and (d) purple paths across the grain boundary. Paths that avoid the grain boundary exhibit relatively small positive MR that changes modestly with temperature, whereas boundary-spanning paths exhibit substantially larger MR at room temperature and evolve toward negative MR upon cooling. These results show that both the MR amplitude and its temperature dependence are strongly path dependent, consistent with enhanced field-sensitive transport at the grain boundary and/or through anisotropic grain-boundary spanning conduction pathways.}
  \label{fig3}
\end{figure}

\noindent  MR, where $\mathrm{MR}(B) = [R(B)-R(0)]/R(B)$, was measured for both the FET and Hall-bar devices (Figure {3}). For the FET devices (Figure 3a), the MR shows monotonic up to 3T.
Large positive MR of 10\% at 3T is observed at room-temperature, whereas large negative MR of approximately -9\% at 3T is observed at
\SI{78.2}{K}. This data exhibits negligible hysteresis. Positive MR at high temperatures typically arises from increased carrier scattering under applied magnetic fields. Upon cooling, the MR becomes negative, indicating trap-limited or
hopping-type transport~\cite{Zhang2021}. In this regime, the magnetic field helps carriers
escape shallow traps and move more easily between localized states, thereby lowering the
resistance. The temperature-dependent resistance analysis (Figure~\ref{fig2}a, inset) is consistent with a crossover from thermally activated transport to hopping conduction, supporting the plausibility of such a transition. A temperature-dependent change in the polarity of MR  has also been observed in other spin–orbit-coupled systems, both in organic materials \cite{Bloom2008} and in inorganic perovskite oxides such as CaRuO$_3$~\cite{Alagoz2024}. Such studies show that competing spin-dependent hopping, localization, and spin–orbit–driven mechanisms could also be responsible for a reversal of the MR sign.  Overall, the magnitude of the low-field MR in our devices ($|\Delta R|/R \sim 10\%$ at $|B|\sim 1$~T) is comparable to that reported for other hybrid lead–halide perovskites~\cite{Wang2019,banerjee2020room,lu2019spin,Yang2020,r10}. 
\\
\vspace{0.01in}
For the Hall-bar sample, the anisotropic behavior observed in the I–V measurements motivated magnetoresistance measurements along multiple current paths (Figure \ref{fig3}b-d). Unlike the three-terminal FET geometry, which was intentionally designed with a wide and short channel to minimize the measurement resistance, the Hall-bar geometry is expected to be approximately 500 times more resistive than the three-terminal channel for the same film thickness and resistivity. Consequently, the resistance measurements on the Hall samples are more susceptible to noise, as is common for low-mobility and high-resistivity samples \cite{Chen2016}. To minimize the impact of noise, we averaged 50 field-sweep cycles from - 5 T to 5 T and 50 field-sweep cycles from 5 T to - 5 T. The Hall-bar MR showed a strong path-dependence consistent with the results of the I-V measurements. While the monotonic and quadratic character of the curves are shared across different paths, measurement paths that traversed across the observed grain boundary exhibit a much larger MR of 6-8\%, comparable with that observed in the FET sample. However, the low resistance current paths that avoided this boundary exhibited a much smaller MR in the range of 0.05-0.2$\%$. The temperature relationship for those current paths traversing the grain boundary follow from the FET measurements, decreasing and becoming negative as temperature decreases (Figure \ref{fig3}d). The current paths that avoid this boundary, however, show a modest increase in MR over the range of temperatures.

The likely origin for this path dependence lies in the fact that the grain boundaries are electronically distinct from transport within a single grain. Grain boundaries in halide perovskites can host sub-bandgap states and band bending.\cite{Qin2021ElectronAccumulationGB,Gunnarsson2002SpinPolarizedGB} In hybrid perovskites, these boundary-associated states can slow carrier diffusion relative to intragrain transport and promote charge accumulation.\cite{Snaider2018UltrafastGB} Such an interfacial region may therefore enhance both the resistance and the magnetic-field sensitivity of current paths that cross the boundary. This interpretation is also consistent with broader grain-boundary magnetotransport literature, where tunneling MR in perovskite manganites is governed by localized defect states and the quality of the interfacial tunneling barrier~\cite{Hoefener2000GB_TMR}.

In layered R-(MBA)$_2$PbI$_4$, this grain-boundary bottleneck may be further amplified by the  transport anisotropy of the organic--inorganic lattice. Current flowing parallel to the inorganic Pb--I sheets is expected to be substantially more conductive than current flowing across the stacked organic--inorganic direction, where carriers must traverse organic spacer barriers. Cheng \textit{et al.} demonstrated this anisotropy directly in related Pb-based hybrid perovskites, showing that current measured parallel to the Pb--I plane was approximately two orders of magnitude larger than current measured perpendicular to the plane; the perpendicular transport direction was further interpreted as tunneling through organic barriers rather than simple Drude transport.\cite{Cheng2018LayerEdge} A grain boundary intersecting the Hall-bar channel can therefore force the current to sample pathways through the organic layer which are not equivalent to in-plane transport which is primarily through the inorganic PbI$_4$ layer. Consequently, the larger longitudinal MR would not necessarily be expected to scale simply with the Hall mobility extracted from a nominally in-plane measurement, because the Hall voltage and longitudinal resistance may weight different portions of the percolating conduction network.
\vspace{0.01 in}
\subsection{3.4 Hall Transport in a Lateral Geometry}
\noindent 
Hall measurements were performed by measuring the transverse voltage $V_{xy}(B)$ for both positive and negative applied current ($+I$ and $-I$) and magnetic-field swept from $-3$T to $+3$T, and $+3$T to $-3$T. Ionic migration is a known complication of transport measurements in lead-iodide perovskites~\cite{Eames2015}. To minimize associated drifts and hysteresis, we employed slow, symmetric field sweeps and allowed the signal to stabilize at each temperature and field setpoint before recording data. The raw data corresponding to these measurements may be found in (Supplementary Fig.~S3). Hall geometries measured at low temperature are sensitive to contact misalignment, slow thermal drifts, and contributions from longitudinal resistance.\begin{wrapfigure}{r}{0.49\textwidth}
\centering
\includegraphics[width=\linewidth]{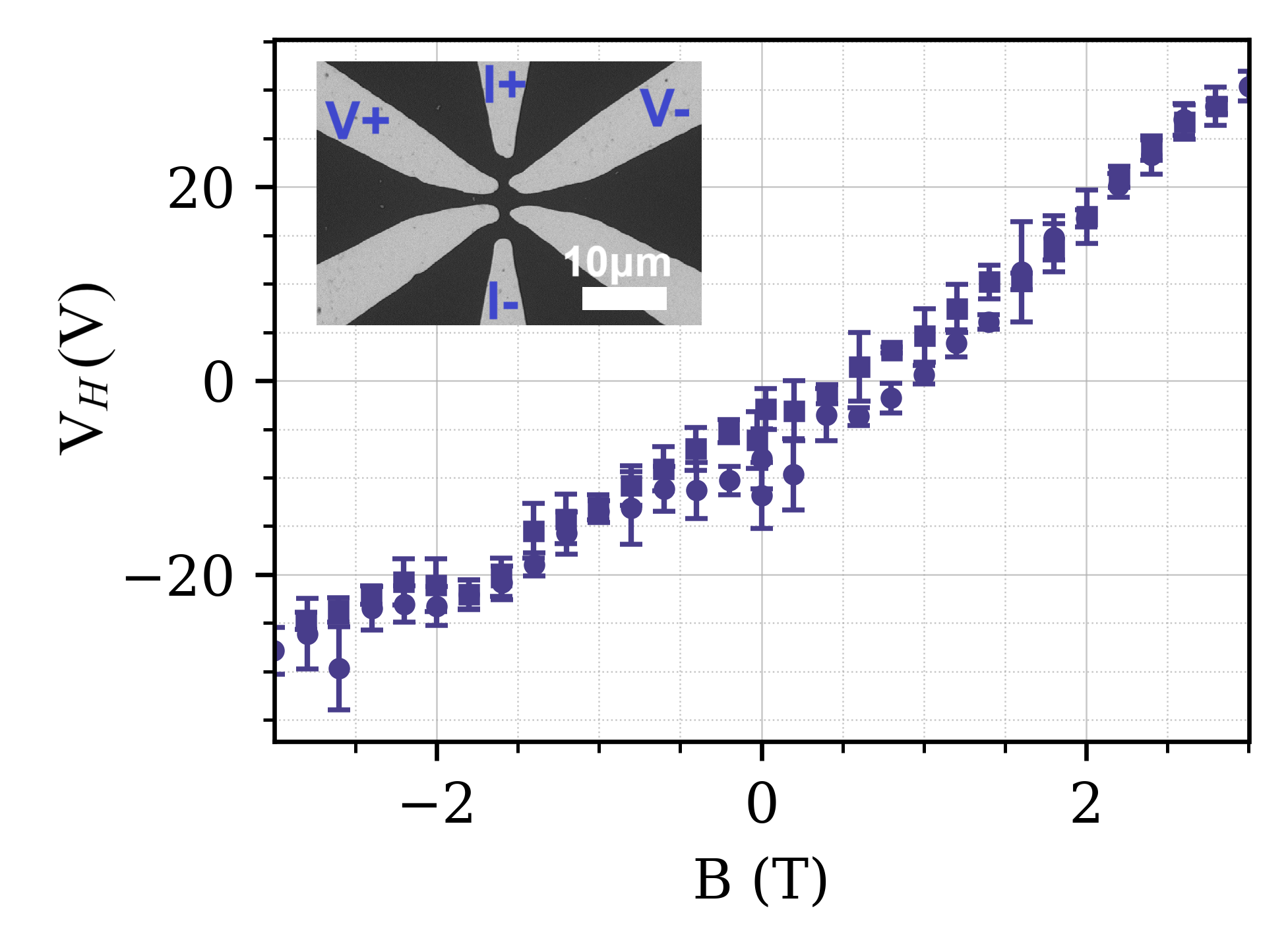}
\caption{Room-temperature Hall effect measurements of R-(MBA)$_2$PbI$_4$ thin films in a Hall-bar geometry. The transverse voltage, $V_{xy}$, was measured between the V$^+$ and V$^-$ contacts while sweeping the magnetic field from $-3$~T to $3$~T. The measured Hall response is consistent with p-type conduction. Inset: SEM image of a representative Hall-bar device showing the measurement geometry.}
\label{fig:hall}
\vspace{-.1 in}
\end{wrapfigure} To suppress signals from these sources, data were first binned in field ($\Delta B = 0.05$~T); within each bin we extracted the component that is odd in current, $V_{xy}^{\mathrm{odd},I}(B) = \frac{1}{2}\big[V_{xy}(B,+I) - V_{xy}(B,-I)\big]$ and then antisymmetrized across $\pm B$ to obtain the Hall voltage, $V_H(B) = \frac{1}{4}\big[V_{xy}(+B,+I) - V_{xy}(+B,-I) - V_{xy}(-B,+I) + V_{xy}(-B,-I)\big]$. Uncertainties in $V_H(B)$ were calculated by using standard error propagation and the measured standard deviation in $V_{xy}$. The resulting Hall voltage (Fig \ref{fig:hall}) $V_H(B)$ varies linearly with $B$ over the full range, and the positive slope of $V_H$ confirms p-type transport in the film. \\
\vspace{0.02 in}
\noindent For a single dominant carrier type, the 2D carrier density is obtained from the Hall coefficient via
$p_{2D} = 1/(q R_H)$, where $q$ is the elementary charge and $R_H$ is extracted from the linear slope of the Hall voltage $V_H(B)$ normalized by the applied current. The Hall mobility is then given by $\mu_H = |R_H|\,\sigma$, where $\sigma$ is the sheet conductivity. At room temperature, the measured Hall coefficient corresponds to a mobility $\mu_H = 0.207$~cm$^2$/V$\cdot$s and a sheet carrier density $p = 3.29 \times 10^{14}$~cm$^{-2}$. This relatively high hole density suggests potential unintentional p-type doping, likely associated with  defect states, and is consistent with prior observations in lead–halide perovskites, where halide vacancies are known to act as shallow acceptors. Comparable Hall mobility values have been reported in related 2D perovskites; for instance, undoped 2D lead-iodide perovskites such as (PEA)$_2$PbI$_4$ typically exhibit mobilities on the order of $\sim$0.1--1~cm$^2$/V$\cdot$s ~\cite{Chen2016}.\\
In a homogeneous classical transport picture, the Hall angle scales as 
$\tan\theta_H \sim \mu B$, while the ordinary orbital magnetoresistance is 
expected to scale approximately as $(\mu B)^2$. Therefore, the low Hall mobility 
extracted here would normally imply a very small classical MR, making the observed 
percent-level room-temperature MR suggestive of an additional transport contribution. We therefore look to the proposed explanation of interface specific transport introduced in the previous section to understand this apparent discrepancy between the measured MR strength and Hall mobility. The Hall voltage reflects an effective averaged response of the conducting network, whereas the longitudinal resistance is likely dominated by the most resistive bottlenecks along the current path. If those bottlenecks correspond to grain-boundary or spacer-mediated transport channels, then they may contribute disproportionately to MR while contributing differently to the Hall response. The magnitude of the observed room-temperature MR is also not unprecedented within the broader hybrid lead-halide perovskite family, where percent-level positive  room-temperature MR approaching 10\% has been reported in out-of-plane MAPbI$_3$ devices.\cite{banerjee2020room} Structurally closer evidence for vertical magnetotransport through layered chiral  lead-iodide HOIPs has also been demonstrated in spin-dependent transport measurements  on two-dimensional chiral hybrid lead iodide perovskites.\cite{lu2019spin} Thus, the present measurements suggest that the large MR in R-(MBA)$_2$PbI$_4$ is  likely not a purely classical orbital effect associated with the Hall mobility, but  instead reflects path-dependent transport through grain-boundary and/or anisotropic cross-layer conduction channels.
\vspace{0.02 in}

\section{Conclusion}
Through the tuning of spin-coating speed—and thereby controlling the thickness and morphology of R-(MBA)$_2$PbI$_4$ films—the in-plane resistivity was reduced by more than three orders of magnitude, enabling reliable lateral device operation.  Consequently, we were able to (i) probe carrier properties in the dark without photo-induced carriers, (ii) perform low-temperature in-plane resistivity and magnetoresistance measurements in FET devices down to 38 K, the lowest temperature reported for in-plane transport in this material,(iii) demonstrate the first direct Hall response and associated transport parameters in this material. Specifically, films spun at 1500 rpm exhibited large, continuous grains with minimal voids and resistance in the M$\Omega$ regime, whereas higher spin speeds yielded thinner, more granular films with resistances in the G$\Omega$ regime. Reduced-resistance films enabled conventional DC measurements in the dark, including gate-tunable field-effect transistor operation, Magnetoresistance, and Hall voltage measurements.
\\

\vspace{0.02 in}
\noindent Both Hall and gate-dependent field-effect measurements indicate that holes are the dominant charge carriers. As the temperature is lowered, strong carrier freeze-out is observed with two distinct regimes: at higher temperatures, transport is governed by  carriers, whereas at lower temperatures it is dominated by carriers released from shallow traps. Qualitatively, this behavior resembles the trends reported in Ref.[24]; however, that study examined quasi-2D mixed-$n$ films rather than a pure $n = 1$ lattice, and these structural differences naturally lead to quantitative deviations. These differences in dimensionality and interlayer coupling naturally lead to quantitative deviations between the two systems.\\

\vspace{0.02 in}
\noindent The material exhibits a small positive MR at room temperature in FET devices, which changes to a negative MR of similar magnitude at 77 K. This sign reversal suggests a temperature-driven change in the dominant charge-transport mechanism underlying charge transport of the material.  In Hall-bar geometries, magnetoresistance exhibits strong transport-path dependence, with enhanced responses observed along transport paths crossing grain boundaries, highlighting the important role of morphology and microstructure in  in-plane transport behavior. Hall voltages were linear with applied field and showed minimal hysteresis, demonstrating little impact from ionic migration. The Hall data were used to calculate the room-temperature Hall carrier concentration, $p = 3.29 \times 10^{14}\,\text{cm}^{-2}$, and the Hall mobility, $\mu_H = 0.207\,\text{cm}^2\,\text{V}^{-1}\,\text{s}^{-1}$. To the best of our knowledge, these measurements represent the first Hall-effect characterization of R-(MBA)$_2$PbI$_4$ thin films under dark conditions.\\ 

\vspace{0.02 in}
\noindent These results provide essential transport parameters for R-(MBA)$_2$PbI$_4$ thin films, including carrier type, carrier density, and the dominant transport mechanisms. The values reported here establish a baseline for room-temperature and low-temperature charge transport in in-plane R-(MBA)$_2$PbI$_4$ devices. Importantly, they also demonstrate that well-controlled R-(MBA)$_2$PbI$_4$ thin films serve as a viable platform for multi-terminal, low-temperature magnetotransport experiments. Given this material’s unique combination of properties—semiconducting behavior, the ability to be deposited as well- oriented layers from solution, and chirality-induced spin selectivity—these advances lay the groundwork for future spin-dependent Hall measurements including insight into the role of grain boundaries in spin transport, and investigations of chirality-induced spin selectivity (CISS).


\section*{Author Information}
\noindent{\large Authors and Affiliations}

\noindent\textbf{Department of Physics, Colorado School of Mines, Golden, CO, USA}\\
Shehreen Aslam, Sam Saiter, Bradley Lloyd, Paul Kliewer, and Meenakshi Singh*

\vspace{1em}

\noindent\textbf{National Laboratory of the Rockies (NLR), Golden, CO, USA}\\
Matthew P. Hautzinger, and Matthew C. Beard
\vspace{1em}

\noindent{\large Author Contributions}\\
\textsuperscript{$\dagger$}Shehreen Aslam and \textsuperscript{$\dagger$}Sam Saiter contributed equally to this work.
\\

\noindent{\large Corresponding author}\\
*Correspondence to Meenakshi Singh

\vspace{2em}

\section*{Acknowledgements}
This work was supported by the U.S. Department of Energy, Office of Science, Basic Energy Sciences, under Award No. SC0024115. We thank the Shared Instrumentation Facility (SIF) at Colorado School of Mines for access to advanced instrumentation and measurement capabilities. Matthew C. Beard and Matthew P. Hautzinger,  acknowledge support from the Center for Hybrid Organic--Inorganic Semiconductors for Energy (CHOISE), an Energy Frontier Research Center funded by the Office of Science, Basic Energy Sciences of the U.S. Department of Energy. This work was authored in part by the National Laboratory of the Rockies (NLR) for the U.S. Department of Energy under Contract No.~DE-AC36-08GO28303. The views expressed in this article do not necessarily represent the views of the U.S. Department of Energy or the U.S. Government.

\vspace{0.02 in}

\bibliography{CISS}
\begin{center}
\vspace*{2in}
{\LARGE \bfseries
Morphology control and low-temperature magnetotransport in chiral 2D perovskite
R-(MBA)$_2$PbI$_4$: Supporting Information
\par}

\vspace{1em}

{\normalsize
\textsuperscript{$\dagger$}Shehreen Aslam,
\textsuperscript{$\dagger$}Sam Saiter,
Bradley Lloyd,
Paul Kliewer,
Matthew P. Hautzinger,
Matthew C. Beard,
and Meenakshi Singh
\par}

\vspace{0.5em}

{\normalsize
\texttt{msingh@mines.edu}
\par}
\end{center}

\vspace{1.5em}
\begin{figure*}[h]
  \centering

  \renewcommand\thesubfigure{\Alph{subfigure}}
  \captionsetup[sub]{labelformat=simple, labelsep=none} 
  \begin{subfigure}[t]{0.32\textwidth}
    \centering
    \includegraphics[width=\linewidth, keepaspectratio]{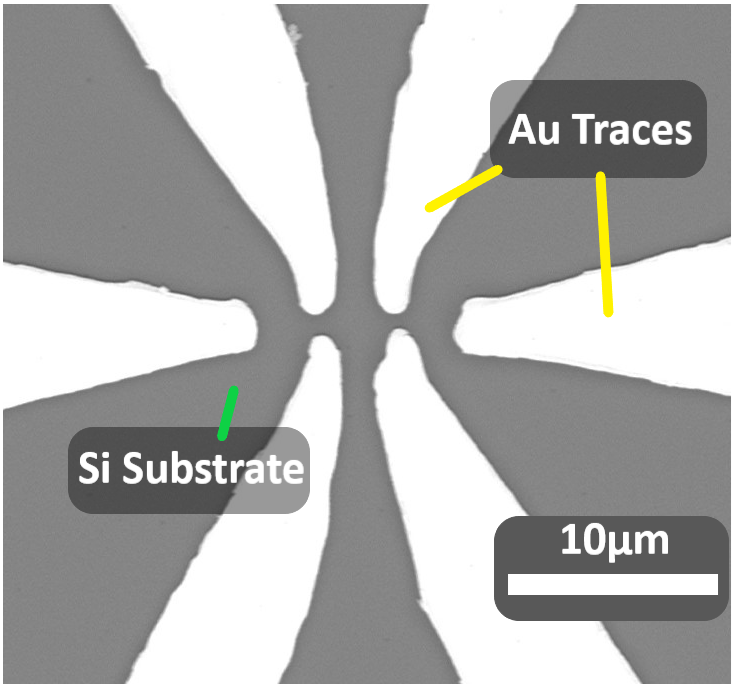}
    \subcaption{}\label{fig:s1-A}
  \end{subfigure}\hfill
  \begin{subfigure}[t]{0.32\textwidth}
    \centering
    \includegraphics[width=\linewidth, keepaspectratio]{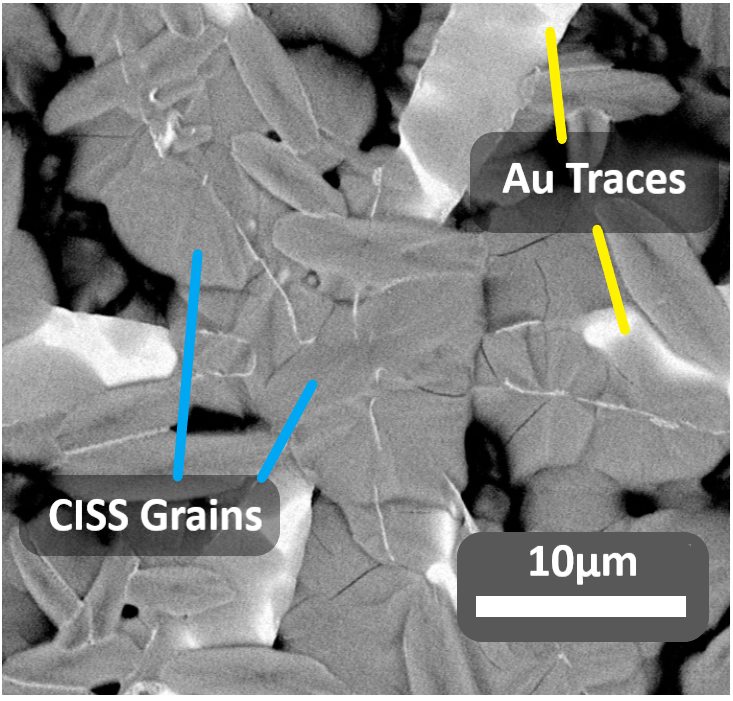}
    \subcaption{}\label{fig:s1-B}
  \end{subfigure}\hfill
  \begin{subfigure}[t]{0.32\textwidth}
    \centering
    \includegraphics[width=\linewidth, keepaspectratio]{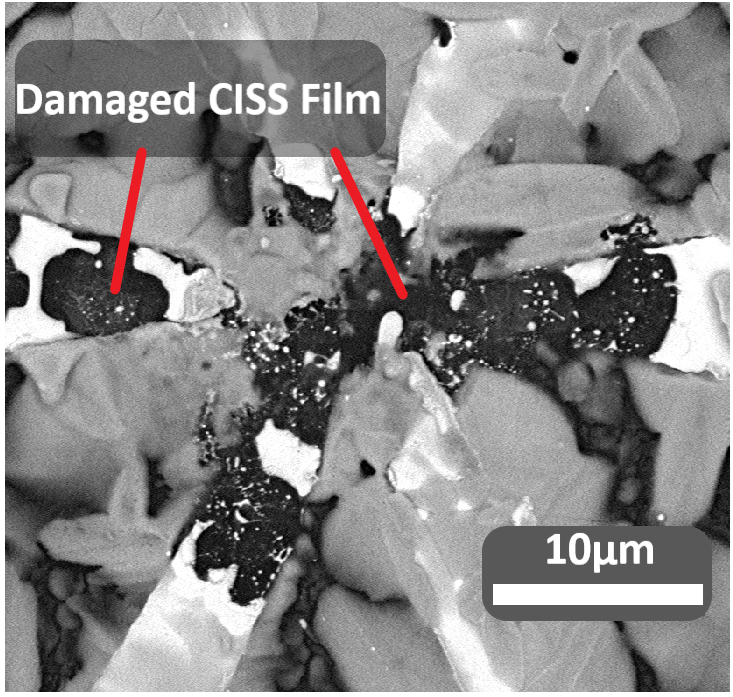}
    \subcaption{}\label{fig:s1-C}
  \end{subfigure}

  \vspace{0.6em}

  \begin{subfigure}[t]{\textwidth}
    \centering
    \includegraphics[width=\textwidth, keepaspectratio]{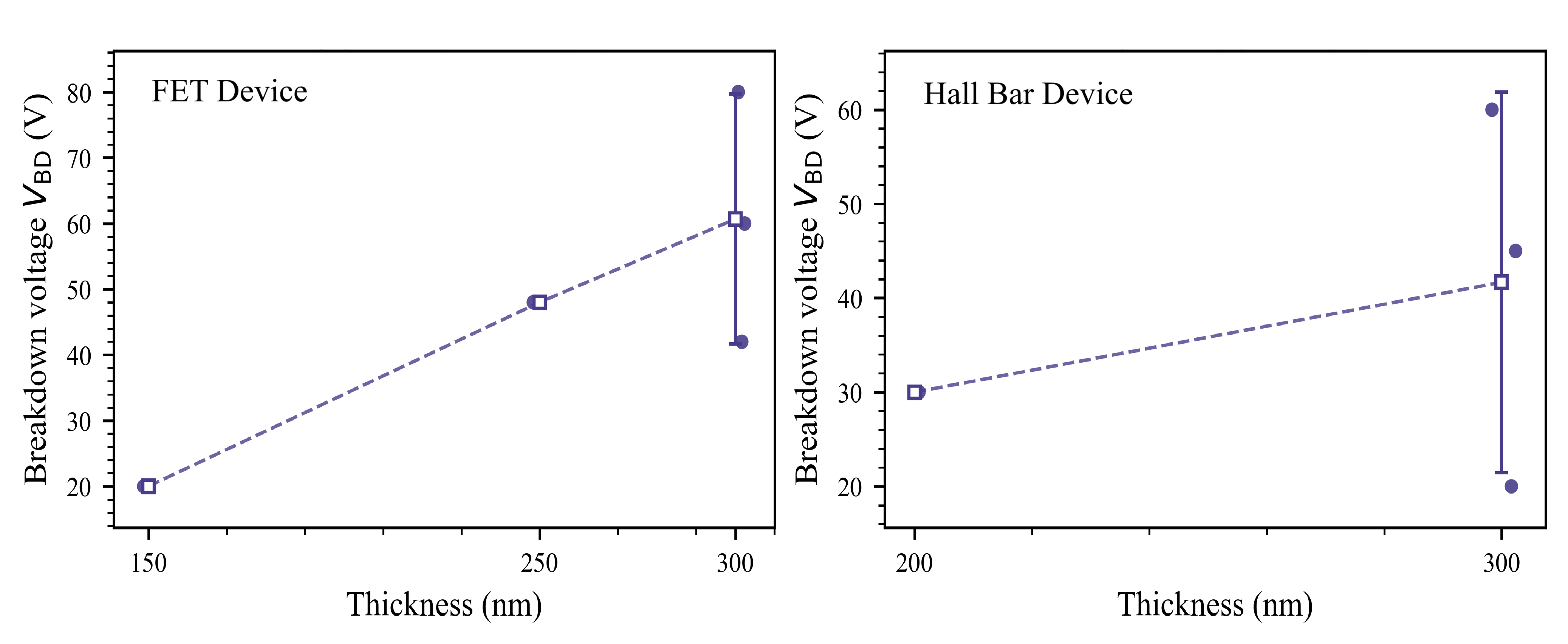}
    \subcaption{}\label{fig:s1-D}
  \end{subfigure}

  \caption{(A) Micrograph of Au contact pads patterned on a SiO$_2$ substrate. (B) Same device after spin-coating the chiral CISS film, \(R\)-(MBA)\(_2\)PbI\(_4\). (C) Example of catastrophic film damage following exposure to large voltages. (D) Breakdown voltage versus film thickness: left, large-area FET-type junctions used in the main text; right, the Hall-bar geometry corresponding to panels A–C.}
  \label{fig:ABCD_composite}
\end{figure*}
\begin{figure}[h]
  \centering
  \includegraphics[width=\linewidth]{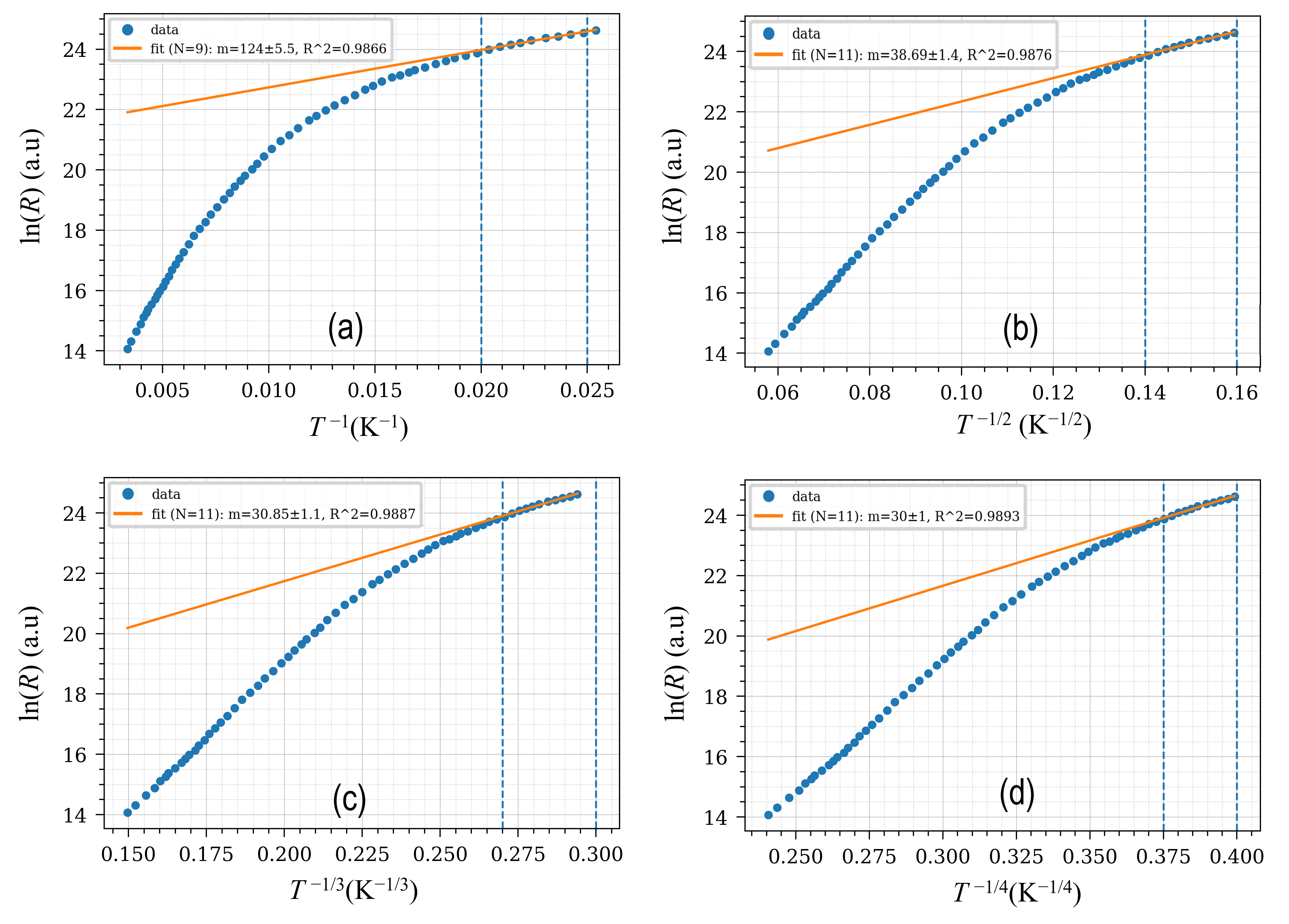}
\caption{$\ln(R)$ plotted versus (a) $T^{-1}$, (b) $T^{-1/2}$, (c) $T^{-1/3}$, and (d) $T^{-1/4}$ to compare Arrhenius and variable-range hopping (VRH) transport models in the low-temperature regime. Linear fits (solid lines) were performed only over the lowest-temperature region, marked by vertical dashed lines, where VRH behavior would be expected to be most prominent. While VRH in low-dimensional semiconductors is typically associated with linear dependence in $\ln(R)$ versus $T^{-1/3}$ (2D) or $T^{-1/4}$ (3D), the present data do not show an obvious improvement in linearization in this regime compared with the Arrhenius plot, and thus do not conclusively distinguish a VRH mechanism from Arrhenius-type transport.}
\end{figure} 
\begin{figure}[h]
  \centering
  \includegraphics[width=0.9\linewidth]{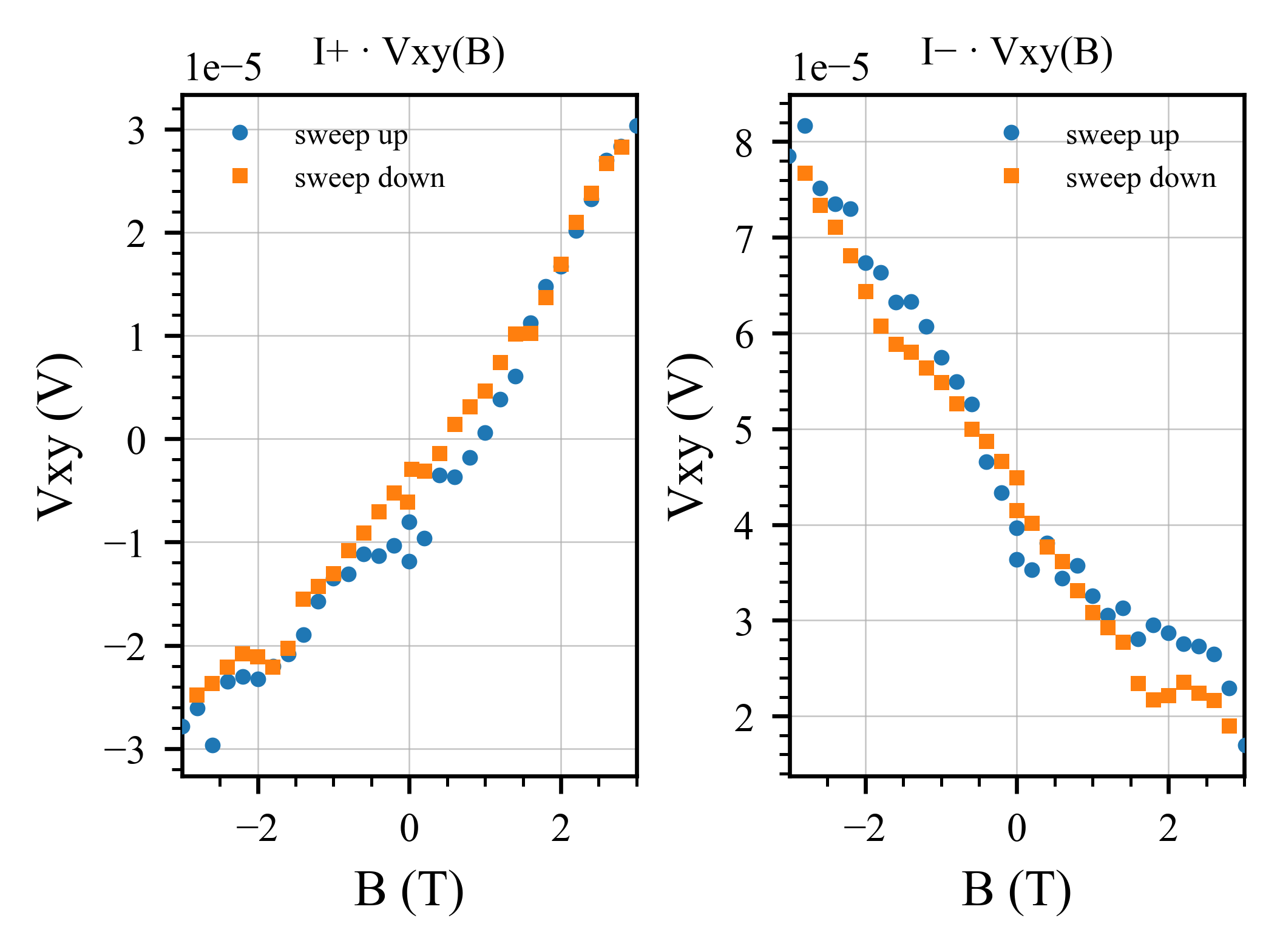}
\caption{Raw Hall voltage $V_{xy}$ versus magnetic field $B$ for $+I$ (left) and $-I$ (right), shown without post-processing over $-3\!\le\!B\!\le\!3$~T. Blue circles denote sweep up and orange squares denote sweep down. The traces exhibit the expected odd-in-field symmetry, $V_{xy}(B)\!\approx\!-V_{xy}(-B)$, with minimal sweep-direction hysteresis. Any small even-in-$B$ background (e.g., contact misalignment or longitudinal pickup) is minor and is removed in the main-text analysis by taking the current-odd component and antisymmetrizing across $\pm B$. This data confirms that the reported Hall response is intrinsic rather than an artifact of signal mixing.}
\end{figure} 

\end{document}